# Atomic-Scale Visualization of the Cuprate Pair Density Wave State

Zengyi Du[1], Hui Li[1,2], and Kazuhiro Fujita*[1]

*[1]Condensed Matter Physics and Materials Science Department, Brookhaven National Laboratory, Upton, NY 11973, USA*
*[2]Department of Physics and Astronomy, Stony Brook University, Stony Brook, NY 11790, USA*

A recent discovery of the pair density wave (PDW) order creates a stir in understanding an underlying physics in the pseudogap states of the cuprate high temperature superconductors. We have performed a Spectroscopic Imaging Scanning Tunneling Microscopy (SI-STM) measurements on $Bi_2Sr_2CaCu_2O_{8+\delta}$ and identified that spatially uniform $d$-wave superconductivity (DSC), $d$-symmetry charge density wave (CDW), and the PDW play important roles in an intertwined fashion. In this review, we show how these different electronic degrees of freedom emerges in electronic structures of the cuprate and discuss how they are related to each other.

## 1. Introduction

**1**    In the $CuO_2$ plane of the mother compound cuprate, each Cu $d_{x^2-y^2}$ orbital is occupied by a single electron and, since the energy required for a double occupancy in this orbital is $U\sim3$ eV, a Mott insulator state develops[1,2]. This situation is well described by the Heisenberg model that is an effective hamiltonian of the Hubbard model, in which the double occupancy is prohibited. The effective spin superexchange interaction $J$ is about $J\sim150$ meV for neighboring $d_{x^2-y^2}$ electron spins, leading to a robust antiferromagnetic (AFM) phase[3,4] (Fig. 1(a)). But this AFM insulating state disappears rapidly upon removing as little as 3% of the electrons per Cu site (hole-density $p$=3%), revealing the pseudogap (PG) state in a region of the phase diagram bounded by $p < p^*$ and temperatures $T < T^*(p)$ (Fig. 1(a)). Key characteristics of the

PG state include[3,4] a rapid decrease in both magnetic susceptibility and $c$-axis (optical) conductivity; an apparently incomplete Fermi surface consisting of relatively coherent quasiparticle states on four momentum space ($\boldsymbol{k}$-space) arcs residing around $\boldsymbol{k} \approx (\pm \pi/2a, \pm \pi/2a)$; an energy gap $\Delta^*$ in the single particle excitation spectrum near $\boldsymbol{k} \approx (\pm \pi/a, 0); (0, \pm \pi/a)$; and the reduction of the average density of states $N(E)$ for $|E| < \Delta^*$ that $\Delta^*(p)$ decreases to zero at $p = p*$ (Fig. 1(a)). An energy-gap in the quasiparticle excitation spectrum in $\boldsymbol{k}$-space occurring only near $\boldsymbol{k} \approx (\pm \pi/a, 0); (0, \pm \pi/a)$ could provide a simple phenomenological explanation for virtually all these PG characteristics, but no comprehensive microscopic theory for the PG phase has yet been established. In this review, we summarize recent studies and discuss a cuprate electronic structure of the puseudogap states based on the PDW perspective[5,6,7].

2.      Historically, a PDW is first proposed as the Fulde-Ferrell-Larkin-Ovchinnikov (FFLO) states (or more precisely as the LO state)[8,9], in which the PDW state is more stable than the spatially uniform SC state when the Fermi surfaces are split by an external magnetic field (Zeeman splitting). In this case, a time-reversal symmetry is broken in the normal state Fermi liquid, and the superconducting susceptibilities $\chi_{sc}(\boldsymbol{q}, T)$ can be peaked at finite $\boldsymbol{q}$. Then, a wavevector of the PDW order parameter is given by $|\boldsymbol{q}| \sim E_z / v_f$, where $E_z$ is the Zeeman energy and $v_f$ is the Fermi velocity, giving rise to a long wavelength of the modulating gap, which is comparable to the coherence length, since the split of the Fermi surface is usually so small[11]. In the case of cuprates, when interactions are not weak, a pair density wave (PDW) state[10,11] has been recently discussed as a leading candidate of the fundamental order parameter that characterizes the pseudogap states. This was originally motivated by transport studies on $La_{2-x}Ba_xCO_4$[12] that led to the hypothetical "stripe superconductivity", in which the superconducting order parameter is spatially modulated and thus a PDW[13,14]. In this case, the PDW order parameter is modulated at 8-unit cell periodicity. Concomitantly, a spin stripe and a charge stripe occur at 8-unit cell and 4-unit-cell periodicities, respectively. The fact is that a variational Monte Carlo study of the striped superconductor, in which double occupancies are removed from the ground state as a result of the strong correlations, revealed that the PDW state is more stable than the spatially uniform SC states or at extremely close competition to each other[18,22]. These results suggest that PDW states can be a mean-field solution for the strongly correlated situations in the absence the external magnetic field. In terms of a single particle excitation spectrum, the highly unusual band structure reconstruction at $T^*$ as observed by the angle resolved photoemission spectroscopy [15] (ARPES) can be explained relatively simply by the formation of a PDW state[16,17]. Indeed, a wide variety of microscopic theories based on strong, local electron-electron interactions now envisage a cuprate PDW state[18,19,20-25], while experimental evidence for its existence is rapidly emerging from multiple techniques[11, 26, 27, 28]. Among them, the scanned Josephson tunneling spectroscopy measurement, in which a spatially modulating Josephson critical current was detected, provides a strong supporting evidence for the PDW to exist[26]. In Fig. 1(b), we show a representative spatially averaged local density of states (LDOS) spectrum of the underdoped $Bi_2Sr_2CaCu_2O_{8+\delta}$, in which a particle-hole symmetric hump (or shoulder) is detected, as denoted by $\Delta_1$. This is where all the pronounced electronic symmetry breakings occur, which we will discuss in this review. $\Delta_0$ in Fig. 1(b) is an energy scale where a Bogoliubov quasiparticle scattering interference disappears, or an energy scale of a possible superconducting gap.

*3.*     Cooper pairs with finite center-of-mass momentum are uniquely characterized by a spatially modulating superconducting energy gap $\Delta(\boldsymbol{r})$[8,9]. Recently, this concept has been expanded to the PDW state predicted to exist in cuprates without an applied magnetic field[10,11]. Although the signature of a cuprate PDW has been detected in Cooper-pair tunneling[26], the definitive signature in single-electron tunneling of a periodic $\Delta(\boldsymbol{r})$ modulation hasn't been observed until recently[5,29]. Thus, the challenge had been to detect the cuprate PDW state using single-electron tunneling. First, we consider a PDW, whose spatial variation of the energy-gap is $\Delta(\boldsymbol{r}) = F_P \Delta_{\boldsymbol{Q}}^P \left[ e^{i\boldsymbol{Q}\cdot\boldsymbol{r}} + e^{-i\boldsymbol{Q}\cdot\boldsymbol{r}} \right]$, where $\Delta_{\boldsymbol{Q}}^P$ is the amplitude of gap modulations at wavevector $\boldsymbol{Q}$, and $F_P$ the form factor with either $s$- or $d$-symmetry. The most obvious and immediate prediction is that the single-electron tunneling should detect a spatially modulating gap at $\boldsymbol{Q}$ in the local density-of-states spectrum $N(\boldsymbol{r}, E)$. It is striking, therefore, that no such modulating $\Delta(\boldsymbol{r})$ has ever been observed in the cuprates. Second, if such a PDW coexists with $d$-wave superconductivity (DSC), whose homogeneous gap is given by $\Delta^S(r) = F_{SC}\Delta^S$, where $F_{SC}$ exhibits $d$-symmetry, then a Ginzburg-Landau (GL) theory predicts the form of $N(r,E)$ modulations induced by the interactions between the PDW $\Delta(r)$ and the superconducting $\Delta^S(r)$. These modulations can be identified from products of these two order parameters that transform as density-like quantities. Thus, considering the product of the PDW and SC order parameters, $\Delta_Q^P\Delta^{S*}$ anticipates $N(r) \propto \cos(\boldsymbol{Q} \cdot \boldsymbol{r})$ modulations at the PDW wavevector $\boldsymbol{Q}$, while the product of a PDW with itself $\Delta_Q^P\Delta_{-Q}^{P*}$ anticipates $N(r) \propto \cos(2\boldsymbol{Q} \cdot \boldsymbol{r})$ modulations at twice the PDW wavevector. Therefore, a second unique signature of a PDW with wavevector $\boldsymbol{Q}$ in the superconducting cuprates would be a coexistence of two sets of $N(r,E)$ modulations at $\boldsymbol{Q}$ and $2\boldsymbol{Q}$. Finally, a topological defect with $2\pi$ phase winding[30] in the induced density wave $N(r) \propto \cos(2\boldsymbol{Q} \cdot \boldsymbol{r})$ is anticipated to induce a local phase winding of $\pi$ in the PDW order, which is designated as a half vortex[31] (Fig. 1(c)). This is the canonical signature of a PDW coexisting with homogeneous superconductivity. Experimental detection of these phenomena in single-electron tunneling constitutes compelling evidence for the PDW state.

## 2. Visualizing the PDW order of the cuprate superconductor

**4** To explore these predictions, we use a spectroscopic imaging scanning tunneling microscopy[32] (SI-STM) with a $Bi_2Sr_2CaCu_2O_{8+\delta}$ nanoflake tip[5] to visualize the single electron tunneling. Utilization of the superconducting tip, if it is created by picking up a flake from the sample *in-situ*, enhances the sensitivity to a gap due to the convolution of spectra that are sharply peaked at superconducting gap edge, in the density-of-states $N_T(E)$ of the tip and $N(r,E)$ of the sample. Similarly, a PDW order parameter in the sample (if any) would be spatially convoluted with the one on the tip when a tunneling junction is formed. Thus, the superconductor-insulator-superconductor (SIS) tunneling technique enhances the sensitivity to probe modulations in $\Delta(r)$. However, it is also the fact that such procedures, by utilizing an unconventional superconducting tip, may involve non-trivial physics behind, and we leave it for a future theoretical study. In this section, a bulk single crystal of $Bi_2Sr_2CaCu_2O_{8+\delta}$ at the hole density $p \sim 0.17 \pm 0.01$ with superconducting transition temperature $T_c$ = 91 K is cleaved at room temperature under ultra-high vacuum condition ($3\times10^{-10}$ Torr), and subsequently inserted into the cryogenic

STM head. The superconducting tip is created by picking up a nanometer scale Bi$_2$Sr$_2$CaCu$_2$O$_{8+\delta}$ flake from the sample[5] to form the SIS junction. The SI-STM measurements throughout this section are then all performed using such SIS junctions at $T$ = 9 K. A typical SIS topography is shown in Fig. 2(a) within a 40 nm × 40 nm field-of-view (FOV). Individual Bi atoms in the BiO plane with sub-atomic resolution are resolved as shown in the inset. The CuO$_2$ plane exists ~ 6 Å below the BiO plane.

**5**     Differential conductance spectra $g(\boldsymbol{r}, E) \equiv dI/dV(\boldsymbol{r}, E = eV)$ at SIS junction are then measured as a function of position in this FOV within ±150 meV. A spatially averaged $g(\boldsymbol{r}, E)$ spectrum is shown in red in Fig. 2(b), together with a normal metal-insulator-superconductor (NIS) spectrum that is obtained on the same sample, but in a different FOV prior to create the superconducting tip. The SIS $g(\boldsymbol{r}, E)$ spectrum, which is a convolution of the tip $N_T(\epsilon)$ and sample $N(\epsilon + E)$ demonstrates an enhanced energy sensitivity as expected (red, Fig. 2(b)). Here, since the spatially averaged NIS $g(\boldsymbol{r}, E)$ spectrum is peaked at ±37 meV while the equivalent SIS spectrum peaks at ±66 meV, the estimated average energy gap of the tip $\Delta_T$ is expected to be 29 meV.

**6**     Next, by measuring half the magnitude of the energy that separates the SIS spectrum peaks at every location, and then subtracting $\Delta_T$, the local gap energy map $\Delta(\boldsymbol{r})$ in the sample is obtained. A typical example of $\Delta(\boldsymbol{r})$ is shown in Fig. 3(a). Figure 3(b) shows the magnitude of the power-spectral-density Fourier transform $\Delta(\boldsymbol{q})$ of $\Delta(\boldsymbol{r})$ from Fig. 3(a). In Fig. 3(b), $\boldsymbol{q}_{SM}$ corresponds to a wavevector of the crystal-structure supermodulation. This supermodulation does indeed produce a type of PDW detectable by its energy gap modulations; but this PDW is a trivial one, occurring due to the modulation of the crystal unit-cell dimensions[5]. Second there is a very broad peak in $\Delta(\boldsymbol{q})$ surrounding $\boldsymbol{q} = 0$ due to the well-known random energy-gap disorder of Bi$_2$Sr$_2$CaCu$_2$O$_{8+\delta}$, and this is equivalent to the widely spread gap values in the histogram shown in the inset of Fig. 3(a). Finally, there are four distinct local maxima in $\Delta(\boldsymbol{q})$ at the points indicated by black solid dots surrounding $\boldsymbol{q} \approx (0, \pm 0.125)2\pi/a_0; (\pm 0.125, 0)2\pi/a_0$.

    These features indicate that there is a strong, if disordered, modulation in $\Delta(\boldsymbol{r})$, running parallel to the Cu-O-Cu bonds of the $CuO_2$ plane. This modulation exists on top of a non-periodic energy gap $\approx 37$ meV. It shows well-defined fourier peaks at $\boldsymbol{Q}_x \approx 2\pi/a_0\,(1/8,0)$ and $\boldsymbol{Q}_y \approx 2\pi/a_0\,(0,1/8)$ indicating that $\Delta(\boldsymbol{r})$ is modulating with $\approx 8a_0$ periodicity along both $x$ and $y$ axes. Such a variation in $\Delta(\boldsymbol{r})$ can be seen directly in a series of SIS $g(\boldsymbol{r},E)$ spectra that are extracted along the line in Fig. 3(a) and are shown in Fig. 3(c). Here we see a local demonstration of how that the energy of maximum $N(\boldsymbol{r})$ (i.e. of the coherence peak) is modulating at $\approx 8a_0$ periodicity in a particle-hole symmetric fashion with an amplitude of approximately 6 meV. More fundamentally, line profiles from $\Delta(\boldsymbol{q})$ in Fig. 3(b) are plotted in Fig. 3(d) for both $x$ and $y$ bond directions. The two well-defined peaks in Fig. 3(d) characterize a PDW with wavevectors $\boldsymbol{Q}_x = 2\pi/a_0(0.129 \pm 0.003 ,0); \boldsymbol{Q}_y = 2\pi/a_0\,(0,0.118 \pm 0.003 )$. This is the first observation of a coherent modulation in the superconducting energy gap $\Delta(\boldsymbol{r})$ expected from the modulating superfluid detected by the Scanned Josephson Tunneling[26], and is precisely what is expected for a PDW state. Moreover, it reveals directly that the cuprate PDW occurs at wavevectors $\boldsymbol{Q} \approx 2\pi/a_0\,(1/8,0); 2\pi/a_0\,(0,1/8)$.

## 3. Symmetry of the density wave induced by the PDW

    As discussed in the previous sections, the product of a PDW order parameter with itself $\Delta_{\boldsymbol{Q}}^P \Delta_{-\boldsymbol{Q}}^{P*}$ predicts the local density states modulations $N(\boldsymbol{r}) \propto \cos(2\boldsymbol{Q} \cdot \boldsymbol{r})$ at twice the PDW wavevector. Thus, one natural consequence is an unconventional charge density wave that modulates the $CuO_2$ Intra-unit-cell (IUC) electronic degree of freedom at $2\boldsymbol{Q}$ since PDW order is expected to have $d$-wave symmetry. Proposals for such exotic CDWs in underdoped cuprates include charge density waves with a $d$-symmetry [33,34,35] and modulated electron-lattice coupling with a $d$-symmetry[36]. Modulations of the IUC states with wavevectors $2\boldsymbol{Q} = (2Q, 2Q); (2Q, -2Q)$ have been extensively studied[37,38,39,40,41] but little experimental evidence for such phenomena has been reported. Most recently, focus has sharpened on the models[42,43] yielding spatial modulations of IUC states that occur at incommensurate wavevectors $2\boldsymbol{Q} = (2Q, 0); (0,2Q)$ aligned with the Cu-O bond directions in the $CuO_2$ plane. The precise mathematical forms of these proposals have important distinctions, and these are discussed in full detail in the ref.6. In this section, we

show site-specific measurements within each $CuO_2$ unit-cell, segregating electronic degree of freedoms into three separate electronic structure images containing only the Cu sites ($Cu(\boldsymbol{r})$) and only the $x/y$-axis O sites ($O_x(\boldsymbol{r})$ and $O_y(\boldsymbol{r})$)[6]. As we shall see below, the phase resolved Fourier analysis reveals directly that the modulations in the $O_x(\boldsymbol{r})$ and $O_y(\boldsymbol{r})$ sublattice images consistently show a relative phase difference of $\pi$. These observations demonstrate, by the direct sublattice phase-resolved visualization, that the charge density wave found in underdoped cuprates consists of modulations of the intra-unit-cell states that exhibit a predominantly $d$-wave symmetry.

**9**    Charge density waves consisting of modulations on the $O_x$ sites that are distinct from those on the $O_y$ sites can be challenging to conceptualize. Therefore, before explaining their modulated versions, we first describe an elementary symmetry decomposition of the IUC states of $CuO_2$. There are three possible configurations[6] since we have three different orbitals in the IUC: a uniform density on the copper atoms with the $O_x$ and $O_y$ sites inactive ($s$-symmetry), a uniform density on the oxygen atoms with copper sites inactive ($s$'-symmetry), and a configuration with opposite-sign of density at $O_x$ and $O_y$ sites being the copper sites inactive ($d$-symmetry). A modulated version of the latter state is shown in Fig. 1(d). As these three IUC arrangements are spatially uniform, they correspond to specific representations of the point group symmetry of the lattice. The Phase-resolved Fourier transforms of each arrangement could reveal their point group symmetry from the relative signs of the Bragg amplitudes. The $s$- and $s$'-symmetry cases both share 90º-rotational symmetry in their Bragg amplitudes, while the Bragg amplitudes for a $d$-symmetry case change sign under 90º rotations[6]. Thus, by measuring the magnitude and sign of the Bragg amplitudes in phase-resolved site-specific electronic structure images, one can extract a degree to which any translational invariant IUC arrangement has an $s$-, $s$'- or $d$-symmetry[6]. Equivalently, we can also judge symmetry from the real space images, $O_x(\boldsymbol{r})$ and $O_y(\boldsymbol{r})$, by taking either sum or subtraction of them. For example, in the case of the $d$ symmetry, if exists, the electronic modulations at $O_x$ and $O_y$ are $\pi$ out of phase, so that $O_x(\boldsymbol{r})+O_y(\boldsymbol{r})$ detects no modulations as they are canceled out, while $O_x(\boldsymbol{r})-O_y(\boldsymbol{r})$ amplifies the modulations. These characteristics can be regarded as an electronic version of the annihilation rule often seen in the X-ray diffraction measurement.

**10**    Next we consider periodic modulations of the IUC states with wavevector $2\boldsymbol{Q}$, which is twice that of the PDW,

$$\rho(\boldsymbol{r}) = [S(\boldsymbol{r}) + S'(\boldsymbol{r}) + D(\boldsymbol{r})]\cos(2\boldsymbol{Q}\cdot\boldsymbol{r} + \phi(\boldsymbol{r})), \qquad (1)$$

that may describe an electronic modulation in the pseudogap states, as shown in Fig. 4A. $\rho(\boldsymbol{r})$ can be a generalized density wave representing whatever electronic degree of freedom is to be modulated, but here we consider CDW; *S, S'* and *D* are the coefficients of the density wave form factors with *s*-, *s'*- and *d*-symmetry, respectively, and $\phi(\boldsymbol{r})$ is an overall phase that can be spatially disordered[6]. A simple way to understand these CDW form factors is to utilize the three $CuO_2$ sublattices: $\boldsymbol{r}_{Cu}$, $\boldsymbol{r}_{O_x}$, $\boldsymbol{r}_{O_y}$ (Fig. 1(d)). By definition, $S(\boldsymbol{r}) = A_S$ for $\boldsymbol{r} \in \boldsymbol{r}_{Cu}$ and otherwise zero; $S'(\boldsymbol{r}) = A_S'$ for $\boldsymbol{r} \in \boldsymbol{r}_{O_x}$ or $\boldsymbol{r}_{O_y}$ and otherwise zero; $D(\boldsymbol{r}) = A_D$ for $\boldsymbol{r} \in \boldsymbol{r}_{O_x}$ ; $D(\boldsymbol{r}) = -A_D$ for $\boldsymbol{r} \in \boldsymbol{r}_{O_y}$ and otherwise zero. The last case is a *d*-symmetry charge density wave (*d*-symmetry CDW) as shown schematically in Fig. 1(d). In cuprates, a generic CDW can have *S, S'*, and *D* all non-zero because the directionality of modulation wavevector $2\boldsymbol{Q}$ breaks rotational symmetry[6]. Therefore, to identify a predominantly *d*-symmetry CDW one needs to consider the real space symmetry operations, which exhibit several distinctive features. Fig. 1(d) shows a schematic *d*-symmetry CDW that modulates along *x* axis. In this state, by considering two trajectories parallel to *x* axis marked as $\phi_x(\boldsymbol{r})$ and $\phi_y(\boldsymbol{r})$, an amplitude of the wave along $O_x$ is exactly $\pi$ out of phase with that along the adjacent trajectory $O_y$. For this reason, electronic modulations are expected to apparently vanish in $O_x(\boldsymbol{r})+O_y(\boldsymbol{r})$ and, when its Fourier transform is determined, no primary modulation peaks occur at ± *2Q* inside the first Brillouin zone (BZ). The second effect is that the Bragg satellite peaks at $\boldsymbol{Q'}$ = $(1,0)±2\boldsymbol{Q}$ and $\boldsymbol{Q''}$=$(0,1)±2\boldsymbol{Q}$ have opposite sign due to a structure factor associated with the site-specific segregation of real space data (see detail in the ref 6). Note, that if an equivalent *d*-symmetry CDW occurred only along $2\boldsymbol{Q}_y$ ($\perp 2\boldsymbol{Q}_x$), then the Bragg satellite peaks occur at $2\boldsymbol{Q'}$ = $(1,0)±2\boldsymbol{Q}_y$ and $2\boldsymbol{Q''}$=$(0,1)±$ $2\boldsymbol{Q}_y$ and again having opposite sign. In Fig. 4(a), we show a typical electronic structure of the pseudogap states for strongly underdoped $Bi_2Sr_2CaCu_2O_{8+\delta}$ (*p*~0.08), visualized by the ratio of the tunneling currents at opposite biases, $R(\boldsymbol{r},E$=150mV)=$I(\boldsymbol{r},+E)/I(\boldsymbol{r},-E)$. We note that the use of $R(\boldsymbol{r},E)$ or

$Z(\mathbf{r}, E) = g(\mathbf{r}, +E)/g(\mathbf{r}, -E)$ is critically important for measuring relative phase of $O_x/O_y$ sites throughout any CDW, because analysis of $g(\mathbf{r}, E)$ shows a systematic error arising from a tip-sample junction formation[32], scrambling the phase information irretrievably. It is also important to note that all the images analyzed below are drift-corrected by the Lawler-Fujita algorithm[44], in which a non-linear distortion of the image due to the scan piezo drift/relaxation is removed. As we shall see below, these procedures exhibit strong modulations of a broken rotational symmetry of the IUC electronic states along both $x$ and $y$ CuO$_2$ bond directions. Fig. 5(a) is a Fourier transform of the image shown in Fig. 4(a), in which $2\mathbf{Q}' = (1,0)\pm2\mathbf{Q}$ and $2\mathbf{Q}''=(0,1)\pm2\mathbf{Q}$ peaks are dominant because of the $d$-symmetry nature of the CDW.

**11**      In Fig 4(b), (c), (d), we show intra-unit-cell electronic degrees of freedom $Cu(\mathbf{r}) \equiv R(\mathbf{r})\delta(\mathbf{r} - \mathbf{r}_{Cu})$, $O_x(\mathbf{r}) \equiv R(\mathbf{r})\delta(\mathbf{r} - \mathbf{r}_{O_x})$, and $O_y(\mathbf{r}) \equiv R(\mathbf{r})\delta(\mathbf{r} - \mathbf{r}_{O_y})$ segregated from Fig. 4(a). While the electronic modulation is not clear in $Cu(\mathbf{r})$ (Fig. 4(b)), $O_x(\mathbf{r})$ and $O_y(\mathbf{r})$ exhibit strong modulations. These situations are further confirmed by the fact that the Fourier Transforms of $O_x(\mathbf{r})$ and $O_y(\mathbf{r})$, as shown in Fig. 5(c), (d), exhibit strong peaks, while the one for $Cu(\mathbf{r})$ does not (Fig. 5(b)). This indicates that the electronic modulations primarily occur at IUC O$_x$ and O$_y$ sites. Now we consider the symmetry operations, and the complex Fourier transforms of $O_x(\mathbf{r})$ and $O_y(\mathbf{r})$, $\tilde{O}_x(\mathbf{q})$ and $\tilde{O}_y(\mathbf{q})$, as shown in Fig. 4(e), (f) and in Fig. 5(e), (f), respectively. Upon calculating the sum $O_x(\mathbf{r}) + O_y(\mathbf{r})$ (Fig. 4(e)), we find no CDW modulations indicating that the modulations in $O_x(\mathbf{r})$ and $O_y(\mathbf{r})$ are canceled to each other, demonstrating that they are indeed $\pi$-out of phase. Equivalently, in $Re\tilde{O}_x(\mathbf{q}) + Re\tilde{O}_y(\mathbf{q})$ (Fig. 5(e)), we find no CDW peaks in the vicinity of $2\mathbf{Q}$. Moreover, there is an evidence for a $\pi$-phase shift between much sharper peaks at $2\mathbf{Q}'$ and $2\mathbf{Q}''$ (albeit with phase disorder). Both of these effects are exactly as expected for a $d$-symmetry CDW (see Fig. 1(d)). Further, upon calculating the subtraction $O_x(\mathbf{r}) - O_y(\mathbf{r})$, we find strong CDW modulations in Fig. 4(f), and the modulation peak at $2\mathbf{Q}$ inside the first BZ, which is absent in Fig. 5(e), is now clearly visible in $Re\tilde{O}_x(\mathbf{q}) - Re\tilde{O}_y(\mathbf{q})$ (Fig. 5(f)). Hence the absence of this feature in $Re\tilde{O}_x(\mathbf{q}) + Re\tilde{O}_y(\mathbf{q})$ cannot be attributed to broadness of the $\mathbf{q}{\sim}0$ features; rather, it is due to a high-fidelity phase cancelation between the modulations on $O_x$ and $O_y$ occurring at $q \sim 2Q$. Finally, the Bragg-satellite peaks at $2Q'=(1,0)\pm 2Q$ and $2Q''=(0,1)\pm 2Q$ that were clear in $Re\tilde{O}_x(q) + Re\tilde{O}_y(q)$ are absent in $Re\tilde{O}_x(q) - Re\tilde{O}_y(q)$. Comparison of all these observations in Fig. 4 and 5 demonstrates that the modulations at $2Q$ maintain a phase difference of approximately $\pi$ between $O_x$ and $O_y$ within each unit cell, and therefore predominantly constitute a $d$-symmetry CDW.

## 4. How the coexisting PDW and superconductor induce CDW modulations

**12**    It has been predicted that a topological defect with $2\pi$ phase winding[30] in the induced density wave $N(r) \propto \cos(2Q \cdot r)$ generates a local phase winding of $\pi$ in the PDW order. This is a new state of matter called the half vortex[31] (Fig. 1(b)). This is the topological signature of a PDW coexisting with homogeneous superconductivity. Experimental detection of these phenomena in single-electron tunneling would constitute compelling evidence for the PDW state. As we saw in the previous sections, the phase-resolved visualization of the $d$-symmetry modulations is applied to the measured $Z(r,E=54\text{meV})$[5] near the gap edge, which is now obtained from the SIS single particle tunneling discussed in the previous section (Fig. 6(a)). A hole density $p$ of this data is $p \sim 0.17$ that slightly differs from the data presented in the previous section. However, the electronic modulation can also be realized in Fig. 6(a). Note that the $Z(r,E=54\text{meV})$ in Fig.6(a) is obtained by the SIS single particle spectroscopy, so it is natural to consider that the electronic modulation seen in Fig. 6(a) is a SIS motif of those discussed in the previous section. Then, we extract the value of $Z(r,E)$ at the oxygen sites within each $CuO_2$ unit cell: $O_x^Z(r) \equiv Z(r)\delta(r - r_{O_x})$ at $O_x$ and similarly for $O_y^Z(r)$ at $O_y$. We then subtract these values throughout the image to yield a map in d-symmetry channel,

$$D^Z(r) = O_x^Z(r) - O_y^Z(r). \qquad (2)$$

Next, we consider the magnitude of the Fourier transform of $D^Z(r,E)$ for $E = 54$ meV where SIS tunneling has the maximum sensitivity to the PDW state[5] (Fig. 2(b)). In the Fourier transform $|D^Z(q, 54 \text{ meV})|$ we find two strong peaks at $Q$ and $2Q$ (Fig. 6(b)) along the line $(0,0) - 2\pi/a_0$ (0.4,0) in Fig. 6(c), from which a Lorentzian background has been subtracted. This density wave structure is the expected signature in $N(r,E)$ modulations[45,46,47] of the PDW with wavevector $Q$ coexisting with the homogeneous superconductivity. Here it is important to note that a peak intensity for $\boldsymbol{Q}$ is much weaker than that for $2\boldsymbol{Q}$ in Fig. 6(c), which is consistent with the theoretical calculations[7].

**13**     Next, by utilizing the two-dimensional lock-in technique[48], the spatial phase $\Phi_{2\boldsymbol{Q}_x}^Z(\boldsymbol{r})$ of $D_{2\boldsymbol{Q}_x}^Z(\boldsymbol{r}, 54\text{ meV})$ is extracted by eq. (5) below,

$$A_{\boldsymbol{Q}}(\boldsymbol{r}) = \int d\boldsymbol{R} A(\boldsymbol{R}) e^{i\boldsymbol{Q}\cdot\boldsymbol{R}} e^{-\frac{(\boldsymbol{r}-\boldsymbol{R})^2}{2\sigma^2}} \qquad (3)$$

$$|A_{\boldsymbol{Q}}(\boldsymbol{r})| = \sqrt{ReA_{\boldsymbol{Q}}(\boldsymbol{r})^2 + ImA_{\boldsymbol{Q}}(\boldsymbol{r})^2} \qquad (4)$$

$$\Phi_{\boldsymbol{Q}}^A(\boldsymbol{r}) = tan^{-1}\frac{ImA_{\boldsymbol{Q}}(\boldsymbol{r})}{ReA_{\boldsymbol{Q}}(\boldsymbol{r})}. \qquad (5)$$

$A(\boldsymbol{r})$ is an arbitrary image, $\boldsymbol{Q}$ is a wavevector of interest to filter out, and $\sigma$ is a coarse graining length. The spatial phase $\Phi_{2\boldsymbol{Q}_x}^Z(\boldsymbol{r})$ that is extracted from eq. (5) is shown in Fig. 7(a). The topological defects with $2\pi$ phase winding in the $N(\boldsymbol{r}, 54\text{ meV})$ density wave are marked by the black and white dots, for which the winding direction is clockwise and counter clockwise, respectively. The presence of these $2\pi$ topological defects in the $N(\boldsymbol{r}, 54\text{ meV})$ density wave at $2\boldsymbol{Q}$, is due microscopically to a dislocation as schematically shown in Fig. 1(b) (black line). To visualize the interplay of the $2\boldsymbol{Q}$ density wave and the PDW, the spatial phase $\Phi_{\boldsymbol{Q}_x}^\Delta$ of the PDW order is extracted in the same way, but now at $2\pi/a_0\ (\pm 1/8, 0)$. It should be noted that a coarse graining length $\sigma$ in Eq. (3) for the Fourier filtration needs to be appropriately chosen. To do this, we extracted $D_{2\mathbb{Q}}(\boldsymbol{r})$ and $\Delta_{\boldsymbol{Q}}(\boldsymbol{r})$ at different cutoff lengths and checked how spatial variations change. Here we chose 16 and 40Å for $D_{2\mathbb{Q}}(\boldsymbol{r})$ and $\Delta_{\boldsymbol{Q}}(\boldsymbol{r})$, respectively, for which spatial variations of $|D_{2\mathbb{Q}}(\boldsymbol{r})|$ and $|\Delta_{\boldsymbol{Q}}(\boldsymbol{r})|$ are insensitive to the cutoff length (see ref.5). In Fig. 7(b), the locations of the $\Phi_{2\boldsymbol{Q}_x}^Z(\boldsymbol{r})$ topological defects from Fig. 7(a) are also plotted on top of the PDW spatial phase $\Phi_{\boldsymbol{Q}_x}^\Delta(\boldsymbol{r})$. Intriguingly, the $\Phi_{2\boldsymbol{Q}_x}^Z(\boldsymbol{r})$ topological defects are always identified in the vicinity of the yellow contours in $\Phi_{\boldsymbol{Q}_x}^\Delta(\boldsymbol{r})$, where the PDW phase is $\pi$. The inset in Fig. 7(b) shows that a distribution of the PDW phase $\Phi_{\boldsymbol{Q}_x}^\Delta$, at which all the topological defects in $\Phi_{2\boldsymbol{Q}_x}^Z(\boldsymbol{r})$ are identified, is clearly centered around $\pi$. Furthermore, an evolution of the PDW spatial phase surrounding the $2\pi$ topological defects along the trajectories in Fig.7(b) shows approximately $\pi$ phase winding in $\Phi_{\boldsymbol{Q}_x}^\Delta$, as shown in Fig.7(c). This is a signature of the possible half-vortex and is expected when a topological defect in the induced density wave at $2\boldsymbol{Q}$ interacts with the PDW order[31].

## 5. Atomic-scale electronic structure of the cuprate PDW state coexisting with superconductivity: a theory-experiment comparison

**14** Finally, we discuss a simple theoretical model to capture STM phenomenology of the cuprate electronic structure[7]. Here, we discuss a strong-coupling mean-field theory of cuprates, in which double occupancies are removed from the ground state, to describe the atomic-scale electronic structure of an eight-unit-cell periodic $d$-symmetry PDW state coexisting with DSC. From this PDW+DSC model, the atomically resolved density of states $N(\boldsymbol{r}, E)$ is calculated at the terminal BiO surface of $Bi_2Sr_2CaCu_2O_{8+\delta}$ and compared with those obtained by the SI-STM. Consistency between theoretical predictions and the corresponding experiments indicates that underdoped hole-doped $Bi_2Sr_2CaCu_2O_{8+\delta}$ does exhibit a PDW+DSC state as presented in the previous sections.

**15** Such a model starts with a Hamiltonian given by $H = -\sum_{\langle i,j \rangle, \sigma} P_G t_{ij} (c_{i\sigma}^\dagger c_{j\sigma} + h.c.) P_G + J \sum_{\langle i,j \rangle} \mathbf{S}_i \cdot \mathbf{S}_j$ , where the $t_{ij}$ describes an electron hopping between Cu $d_{x^2-y^2}$ orbitals at $i$ and $j$ sites, $J$ is an antiferromagnetic superexchange interaction, and the $P_G$ is a Gutzwiller projection operator that removes all doubly occupancies in the states from the Hilbert space[7]. A renormalized mean-field theory (RMFT) approximation applied to this $t$-$J$ model replaces the PG with renormalization factors $g_i^t$ and $g_i^s$ that are determined by the average number of charge and spin configurations at every Cu site[49]. Then, the Hamiltonian is mean-fields-decoupled into a simpler but diagonalizable form, in which the mean-fields are on-site hole density $\delta_i$, bond field $\chi_{ij\sigma}$, and electron-pair field $\Delta_{ij\sigma}$[7]. Subsequently, a set of Bogoliubov-de Gennes (BdG) equations, together with self-consistency equations for the mean fields are obtained by variational minimization of the ground state energy with respect to the unprojected wavefunction $|\Psi_0\rangle$. Translational symmetry breaking within RMFT is introduced by having site-specific and bond-specific renormalization factors $g_{i,j}^t$ and $g_{i,j}^s$ for charge and spin[7]. To get a PDW+ DSC solution, the BdG equations are initialized with a set of order parameter fields modulating at wavevector $\boldsymbol{Q_P} = (1/8,0)2\pi/a_0$, and a self-consistent solution is obtained. Consequently, various mean fields such as the net charge on each Cu site, the bond field between adjacent Cu sites $i$, $j$, and the electron-pair field on the bond between adjacent Cu sites I, j are calculated, respectively, from the consequent many-body wavefunction $\Psi_0(\boldsymbol{r})$ of this broken-symmetry state;

$$\delta_i = 1 - \langle \Psi_0 | \sum_\sigma n_{i\sigma} | \Psi_0 \rangle \tag{6}$$

$$\chi_{ij\sigma} = \langle \Psi_0 | c_{i\sigma}^\dagger c_{j\sigma} | \Psi_0 \rangle \tag{7}$$

$$\Delta_{ij\sigma} = \sigma \langle \Psi_0 | c_{i\sigma} c_{j\bar{\sigma}} | \Psi_0 \rangle . \tag{8}$$

It should be noted that the commensurate PDW+DSC studied here is not the ground state of the cuprate RMFT Hamiltonian, but that its energy above the homogeneous ground state is so tiny[23,24],[50] (~1meV see ref.7) that it can be stabilized by a variety of means, including disorder.

**16**    Using this model, the atomic-scale characteristics of a unidirectional, eight-unit-cell periodic PDW state coexisting with uniform DSC are examined. For comparison with underdoped $Bi_2Sr_2CaCu_2O_{8+\delta}$ measurements, quasiparticle states with intra-unit-cell resolution, which is obtained by the RMFT PDW+DSC model with a Wannier function-based method are evaluated[24],[51]. This allows quantitative predictions of electronic structure in $\boldsymbol{r}$-space, $\boldsymbol{q}$-space and $\boldsymbol{k}$-space of a PDW+DSC state. The band structure parameterization is $t$=400meV, $t' = -0.3t$ and $J=0.3t$, all representing $Bi_2Sr_2CaCu_2O_{8+\delta}$ at $p \approx 0.08$ . For this parameter set, the BdG equations are solved self-consistently and the lattice Green's function is calculated for the PDW+DSC state, with the unidirectional PDW wavevector $\boldsymbol{Q_P} = (1/8,0)2\pi/a_0$ that is modulating parallel to the $x$-axis but lattice-periodic along the $y$-axis, and find the PDW spectral gap $\Delta_1 \approx 0.3t \approx 100$meV, and gap for uniform DSC $\Delta_0 \approx 0.07t \approx 25$meV[7]. Moreover, while all previous RMFT studies of cuprates yield only the Cu site-specific Green's function $G_{ij\sigma}(E)$ within the $CuO_2$ plane, the experimental measurements of electron-tunneling are performed at a continuum of locations just above the crystal terminal BiO layer of $Bi_2Sr_2CaCu_2O_{8+\delta}$. Therefore, the lattice Green's function is transformed into a continuum Green's function by using the Wannier functions $W_i(\boldsymbol{r})$ obtained by the first-principles calculations for Cu $d_{x^2-y^2}$ orbital; the $\boldsymbol{r}$-space continuum

Green's functions $G_\sigma(\boldsymbol{r}, E) = \sum_{ij} G_{ij\sigma}(E) W_i(\boldsymbol{r}) W_j^*(\boldsymbol{r})$ of a PDW+DSC state, at a height 0.4 nm above BiO terminal plane[52,53]. Note that none of the mean fields $\delta_i$, $\chi_{ij\sigma}$ and $\Delta_{ij\sigma}$ are related simply to the local quasiparticle density of states $N(\boldsymbol{r}, E) = -\sum_\sigma \frac{1}{\pi} \operatorname{Im} G_\sigma(\boldsymbol{r}, E)$, which must instead be determined from the Bogoliubov quasiparticle eigenstates[54,55,56] that enter the lattice Green's function.

**17**     The bias dependence of $Z(\boldsymbol{r}, E) = N(\boldsymbol{r}, +E) / N(\boldsymbol{r}, -E)$ from the PDW+DSC model are then calculated for a comparison with the experiment. Figs. 8(a-f) show these $Z(\boldsymbol{r}, E)$ data from the energy range $0.5 \, \Delta_1 \lesssim E \lesssim 1.5 \, \Delta_1$, where a $d$-symmetry charge modulation is realized[48]. This effect can be seen directly because the model $Z(\boldsymbol{r}, E)$ has intra-unit-cell precision. Here, again, the three sublattices electronic degree of freedom in $Z(\boldsymbol{r}, E)$ are considered: $Cu(\boldsymbol{r}, E)$ containing only $Z(\boldsymbol{r}, E)$ at copper sites and $O_x(\boldsymbol{r}, E)$ and $O_y(\boldsymbol{r}, E)$, containing only $Z(\boldsymbol{r}, E)$ at the $x/y$-axis oxygen sites. By definition, in a $d$-symmetry CDW, modulations on the $O_x(\boldsymbol{r}, E)$ and $O_y(\boldsymbol{r}, E)$ sites are $\pi$ out of phase. In the PDW+DSC model, such phenomena occur at both $\boldsymbol{Q_P}$ and $2\boldsymbol{Q_P}$, first appear near $E \approx \Delta_1/2$, are intense surrounding $E \approx \Delta_1$ and eventually disappear near $E \approx 2\,\Delta_1$[7].

**18**     Figures. 8(g-l) show the measured $Z(\boldsymbol{r}, E)$ in the energy range $0.5\Delta_1 \lesssim E \lesssim 1.5\Delta_1$, with each panel compared side-by-side with the equivalent energy from the model. The most intense modulations occur at $E \approx \Delta_1(p)$ for all $p < 0.19$[48,57], and through these energies range they exhibit a predominant $d$-symmetry CDW[6,48,57]. Therefore, correspondence between theoretical $Z(\boldsymbol{r}, E)$ from the PDW+DSC model (Fig. 8(a-f) ) and the measured $Z(\boldsymbol{r}, E)$ (Fig. 8(g-l)) appears to be in excellent agreement over a wide energy range. In fact, a cross-correlation value of each pair of theory-experiment $Z(\boldsymbol{r}, E/\Delta_1)$ images is shown in Fig. 8(m), exhibiting strong similarities between theory-experiment pairs of $Z(\boldsymbol{r}, E)$ images throughout the energy range. Thus, predictions of the PDW+DSC theory, at distance scales ranging from eight-unit-cell down to sub-unit cell, highly correspond to the electronic structure observed in the underdoped regime in $Bi_2Sr_2CaCu_2O_{8+\delta}$.

**19**      Next, the theory predicts that coherence peak energy $\Delta_1$ varies substantially from one unit cell to the next within the eight-unit-cell periodicity of the PDW. The theoretical gap map and its linecut along $x$ direction for the PDW+DSC state are shown in Fig. 9(a) and (c). The gap map is obtained by extracting the peak energy $\Delta_1(\boldsymbol{r})$ for $\omega > 0$ at all intra-unit cell points over an area of $8\times12$ unit-cells. Similarly, the gap map obtained by using the same algorithm to determine $\Delta_1(\boldsymbol{r})$ from measured $dI/dV$ spectra and its linecut along the $x$-axis (averaged along $y$-axis) are shown in Fig. 9(b) and (d). Both the theory and experiment show eight-unit-cell periodic $\Delta_1(\boldsymbol{r})$ modulations that there are smaller atomically resolved variations exhibiting common characteristics but not identical, most likely due to inadequacies in the DFT-derived Wannier orbitals in representing underdoped cuprates.

## 6. Summary

**20**      To recapitulate, it is demonstrated that the electronic structure near the pseudogap energy scale is characterized by the modulating $\Delta_1(\boldsymbol{r})$ at eight-unit-cell periodicity of the PDW as well as the modulating LDOS primarily at four-unit-cell periodicity of the CDW. The CDW is predominantly $d$-symmetry, for which the IUC electronic degrees of freedom at $O_x$ and $O_y$ sites are modulated out-of-phase by $\pi$. It is also demonstrated that such four-unit-cell $d$-symmetry CDW is naturally induced by the presence of the PDW order as a result of the coupling of the PDW itself. The modulating LDOS at eight-unit-cell periodicity is rather weak in NIS, but detected clearly in SIS single particle tunneling spectroscopy. The observation of this modulation is a supporting evidence for the PDW coupled to DSC order parameter. The proposed strong-coupling mean-field theory for a coexisting DSC and PDW is compared in detail with the experimental SI-STM data for $Bi_2Sr_2CaCu_2O_{8+\delta}$. Indeed, the PDW+DSC model explains simply the microscopic origins of many enigmatic characteristics of this broken-symmetry state in the $Bi_2Sr_2CaCu_2O_{8+\delta}$ cuprate superconductors. It was also demonstrated that the $2\pi$ topological defects in $d$-symmetry CDW generated the $\pi$ phase winding in the PDW order parameter, which is a predicted characteristic of the half-vortex, indicating that the PDW plays a fundamental role in the cuprate playground.

**Acknowledgement**

This work is done in collaboration with Andrea Allais, Masaki Azuma, Peayush Choubey, Stephen D. Edkins, J. C. Séamus Davis, Elizabeth. P. Donoway, Hiroshi Eisaki, Genda Gu, Mohammad H. Hamidian, Peter. J. Hirschfeld, Peter D. Johnson, Sanghyun Joo, Chung Koo Kim, Eun-Ah Kim, Yuhki Kohsaka, Michael J. Lawler, Jinho Lee, Andrew P. Mackenzie, Abhay N. Pasupathy, Subir Sachdev, Hidenori Takagi, Mikio Takano, and Shin-ichi Uchida. We thank E. Fradkin, S. A. Kivelson, Y. S. Lee, P. A. Lee, and J. M. Tranquada, for fruitful discussions and excellent advice. Z.D., H.L. and K.F. acknowledge support from the U.S. Department of Energy, Office of Basic Energy Sciences, under contract number DEAC02-98CH10886.

*e-mail: kfujita@bnl.gov

**Figure captions**

**Fig. 1.** (Color) (a) A schematic phase diagram of hole-doped cuprates. $T^*$ marks boundary of the pesudogap (PG) phase. $T_{sc}$ represents the critical temperature of superconductivity. dCDW is the d-symmetry charge density wave states. (b) A typical differential conductance spectrum measured in the underdoped cuprate. Dash lines identify two characteristic energies of $\Delta_0$ and $\Delta_1$. (c) A PDW state intertwined with induced density wave modulations in $N(\boldsymbol{r})$ in the presence of an edge dislocation in the induced density wave. The unidirectional PDW modulations and the induced density wave modulations in $N(\mathbf{r})$ have $8a_0$ and $4a_0$ periodicity, respectively (color plot and black broken lines, respectively). Around the dislocation in the induced density wave, $N(\boldsymbol{r})$ tends to have $2\pi$ phase winding, while PDW states have $\pi$ phase winding (half-vortex). (d) Schematic of the electronic structure for the d-symmetry CDW. Grey dots represent the Cu sites, and the $O_x$ and $O_y$ sites within each $CuO_2$ unit-cell are electronically inequivalent. The schematic CDW

modulates horizontally with wavelength $\lambda$, or with wavevector $2\boldsymbol{Q}$, and with period $4a_0$. The periodic modulations at $O_x$ sites are $\pi$ out of phase with those at $O_y$ sites, as seen by considering the two trajectories marked by $\phi_x$ and $\phi_y$. (b) is reproduced from *Proc. Natl' Acad. Sci.* **117** 14805 (2020) National Academy of Sciences. (c) is reproduced from *Nature* **580**, 65 (2020) Springer Nature. (d) is reproduced from *Nature Phys.* **12**, 150(2015) Springer Nature.

**Fig. 2.** (Color) (a) Typical SIS topography of $Bi_2Sr_2CaCu_2O_{8+\delta}$ within 40 nm × 40 nm FOV. Cu-O-Cu bond directions are parallel to the $x$ and $y$ axes. Individual Bi atoms are clearly observed as shown in the inset with intra-unit-cell resolution. (b) Spatially averaged SIS $g(E)$ spectrum is shown in red, together with that taken with NIS junction in black. This figure is reproduced from *Nature* **580**, 65 (2020) Springer Nature.

**Fig. 3.** (Color) (a) Measured $\Delta(\mathbf{r})$ within FOV in Fig. 2A by SIS single particle tunneling. The inset shows a distribution of measured $\Delta(\mathbf{r})$ in the same FOV. (b) A magnitude Fourier transform of (a). (c) Intensity plot of the series of SIS $g(\mathbf{r}, E)$ spectra along the line in (a). (d) A line cut of $|\Delta(\mathbf{q})|$ along both $x$ and $y$ directions from (b). This figure is reproduced from *Nature* **580**, 65 (2020) Springer Nature.

**Fig.4.** (Color) (a) Measured $R(\mathbf{r},150\text{ meV})$ for $Bi_2Sr_2CaCu_2O_{8+\delta}$ sample with $p{\sim}8\pm1\%$. (b) Electronic degree of freedom at copper site, $Cu(\mathbf{r})$ extracted from (a), in which the spatial average is subtracted. (c) Same as (b), but at $x$-bond oxygen-site, $O_x(\mathbf{r})$, showing strong modulation. (d) Same as (b), but at y-bond oxygen-site, $O_y(\mathbf{r})$, showing strong modulation. (e) $O_x(\mathbf{r}) + O_y(\mathbf{r})$, showing no modulation. (f) $O_x(\mathbf{r}) - O_y(\mathbf{r})$, showing strong modulation. This figure is reproduced from *Proc. Natl' Acad. Sci.* **111**, E3026 (2014) National Academy of Sciences.

**Fig. 5.** (Color) (a) Magnitude Fourier transform $R(\mathbf{q}, 150\text{ meV})$ of $R(\mathbf{r}, 150\text{ meV})$. The CDW peaks appear at about three quarters as Bragg-satellite peaks, not at $2\boldsymbol{Q}$. (b) $\text{Re}\widetilde{Cu}(\boldsymbol{q})$, a real part of the Fourier transform of $Cu(\mathbf{r})$ in Fig. 4b. No CDW peaks are found. (c)

$\mathrm{Re}\tilde{O}_x(\boldsymbol{q})$, a real part of the Fourier transform of $O_x(\boldsymbol{r})$ in Fig. 4c. The four CDW peaks at $2\boldsymbol{Q}$ and the CDW Bragg-satellite peaks are resolved. (d) $\mathrm{Re}\tilde{O}_y(\boldsymbol{q})$, a real part of the Fourier transform of $O_y(\boldsymbol{r})$ in Fig. 4d. Again, the four CDW peaks at $2\boldsymbol{Q}$ and the CDW Bragg-satellite peaks are resolved. (e) $\mathrm{Re}\tilde{O}_x(\boldsymbol{q}) + \mathrm{Re}\tilde{O}_x(\boldsymbol{q})$. Equivalently, the Fourier transform of Fig. 4e. The four CDW peaks at $2\boldsymbol{Q}$ are not detectable whereas the CDW Bragg-satellite peaks are enhanced and well resolved. (f) $\mathrm{Re}\tilde{O}_x(\boldsymbol{q}) - \mathrm{Re}\tilde{O}_x(\boldsymbol{q})$. Equivalently, the Fourier transform of Fig. 4f. The primary CDW peaks at $2\boldsymbol{Q}$ are strongly enhanced whereas the CDW Bragg-satellite peaks disappeared. This figure is reproduced from *Proc. Natl' Acad. Sci.* **111**, E3026 (2014) National Academy of Sciences.

**Fig. 6.** (Color) (a) $Z(\boldsymbol{r}, 54\ \mathrm{meV})$ around the pesudogap energy. (b) A magnitude of the phase-resolved Fourier transform, $|D^Z(\boldsymbol{q})|$ exhibiting both $\boldsymbol{Q} = 2\pi/a_0(\pm 1/8, 0)$ and $2\boldsymbol{Q} = 2\pi/a_0(\pm 1/4, 0)$ peaks indicated by red broken lines, respectively. (c) A line cut of $|D^Z(\boldsymbol{q})|$, in which Lorentzian background is subtracted, in (b), exhibiting well-defined peaks at $\boldsymbol{Q}$ and $2\boldsymbol{Q}$. The solid line is a curve obtained by Lorentzians fitting. Obtained peak positions are $2\pi/a_0(0.113 \pm 0.002)$ and $2\pi/a_0(0.241 \pm 0.003)$ for $\boldsymbol{Q}$ and $2\boldsymbol{Q}$, respectively with the peak widths to be $2\pi/a_0(0.016 \pm 0.004)$ and $2\pi/a_0(0.068 \pm 0.006)$ for $\boldsymbol{Q}$ and $2\boldsymbol{Q}$, respectively. This figure is reproduced from *Nature* **580**, 65 (2020) Springer Nature.

**Fig. 7.** (Color) (a) $\Phi^Z_{2\boldsymbol{Q}_x}(\boldsymbol{r})$, a spatial phase of $2\boldsymbol{Q}$ modulation in $N(\boldsymbol{r})$ obtained by eq. (3,5). $2\pi$ topological defects are marked by solid dots. White (black) dots indicate $2\pi$ phase winding along clockwise (counterclockwise) direction. (b) $\Phi^\Delta_{\boldsymbol{Q}_x}(\boldsymbol{r})$, a spatial phase of $\boldsymbol{Q}$ modulation in $\Delta(\boldsymbol{r})$. The $2\pi$ topological defects in $\Phi^Z_{2\boldsymbol{Q}_x}(\boldsymbol{r})$ from (a) are plotted on top of $\Phi^\Delta_{\boldsymbol{Q}_x}(\boldsymbol{r})$. The inset shows a distribution of $\Phi^\Delta_{\boldsymbol{Q}_x}(\boldsymbol{r})$ values at all the locations where the $2\pi$ topological defects in $\Phi^Z_{2\boldsymbol{Q}_x}(\boldsymbol{r})$ are found. (c) An evolution of the phase shift along each contour in (b) that encircles the $2\pi$ topological defects found in $\Phi^Z_{2\boldsymbol{Q}_x}(\boldsymbol{r})$. Symbols represent a corresponding start (end) point of the contour in (b). This figure is reproduced from *Nature* **580**, 65 (2020) Springer Nature.

**Fig. 8.** (Color) (a–f) Predicted $Z(r,E)$ maps in PDW + DSC state for the terminal BiO layer, at energies $|E| = 0.4\Delta_1$, $0.6\Delta_1$, $0.8\Delta_1$, $\Delta_1$, $1.2\Delta_1$, and $1.4\Delta_1$, respectively. (g–l) Experimental $Z(r,E)$ maps for bias voltages corresponding to energies in (a–f), respectively. (m) Cross correlation coefficients between theoretical and experimental $Z(r,E)$ maps at different energies, exhibiting very strong correspondence for an energy range $0.5\Delta_1$-$1.5\Delta_1$ (marked with vertical black dashed lines) around PDW energy gap scale. This figure is reproduced from *Proc. Natl' Acad. Sci.* **117** 14805 (2020) National Academy of Sciences.
.

**Fig. 9.** (Color) (a) Gap map $\Delta_p(r)$ derived from coherence-peak energy in PDW + DSC state. Color bar is given in the units of $t$. (b) Gap map $\Delta_p(r)$ obtained by the same procedure as in (a), but from experimental $g(r,V)$ spectra of a representative domain. (c) A spatial evolution of the gap $\Delta_p(r)$ along $x$ axis obtained from the PDW + DSC model. (d) A spatial evolution of the experimental gap $\Delta_p(r)$ along $x$ axis, which is averaged along $y$ direction. This figure is reproduced from *Proc. Natl' Acad. Sci.* **117** 14805 (2020) National Academy of Sciences.

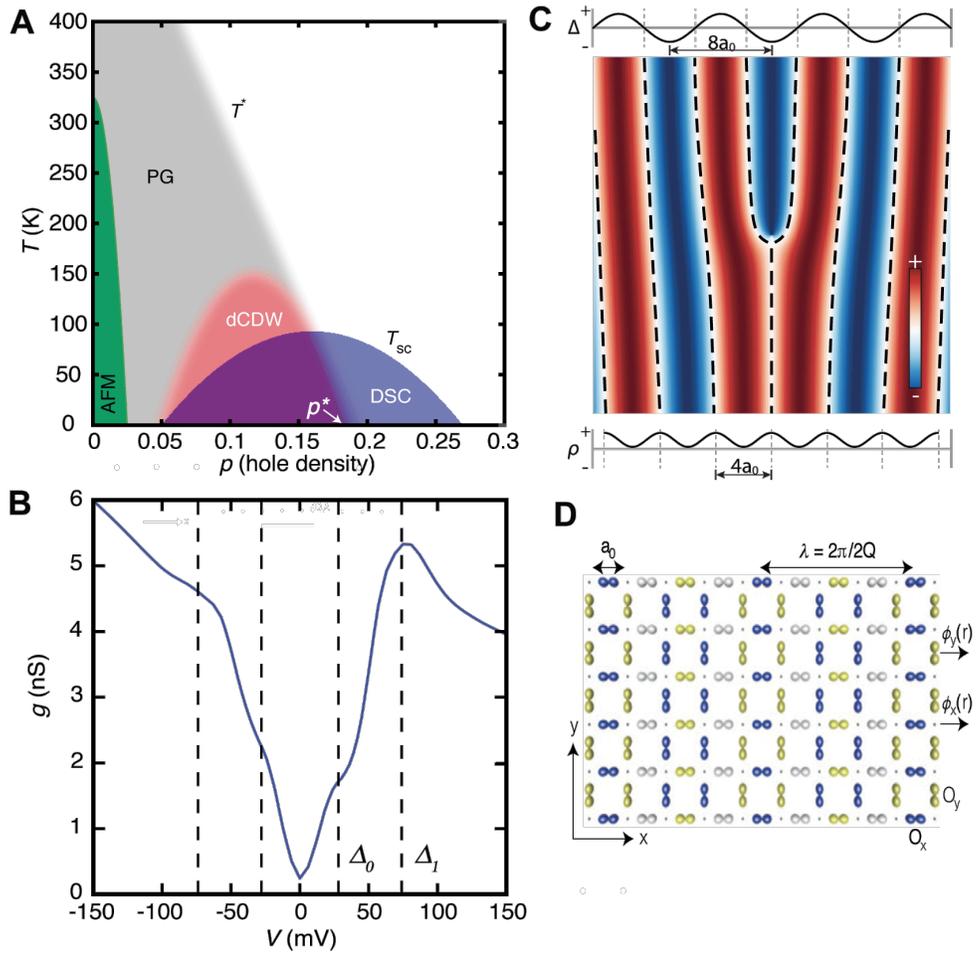

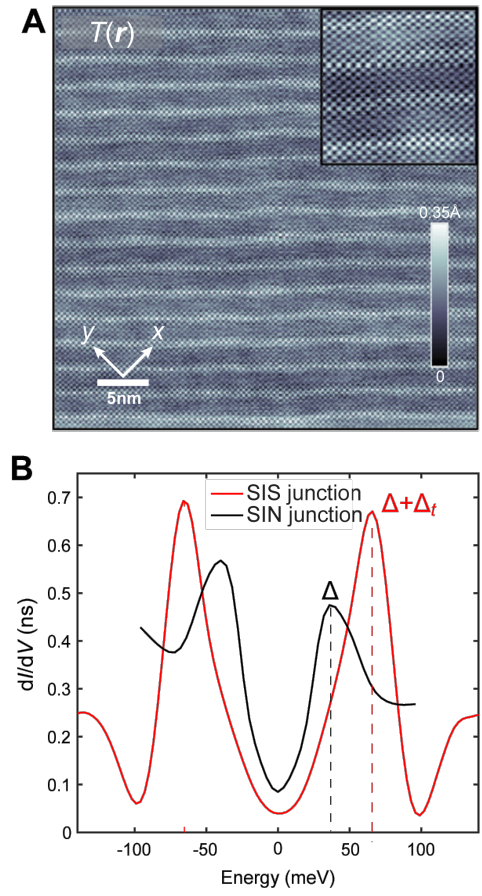

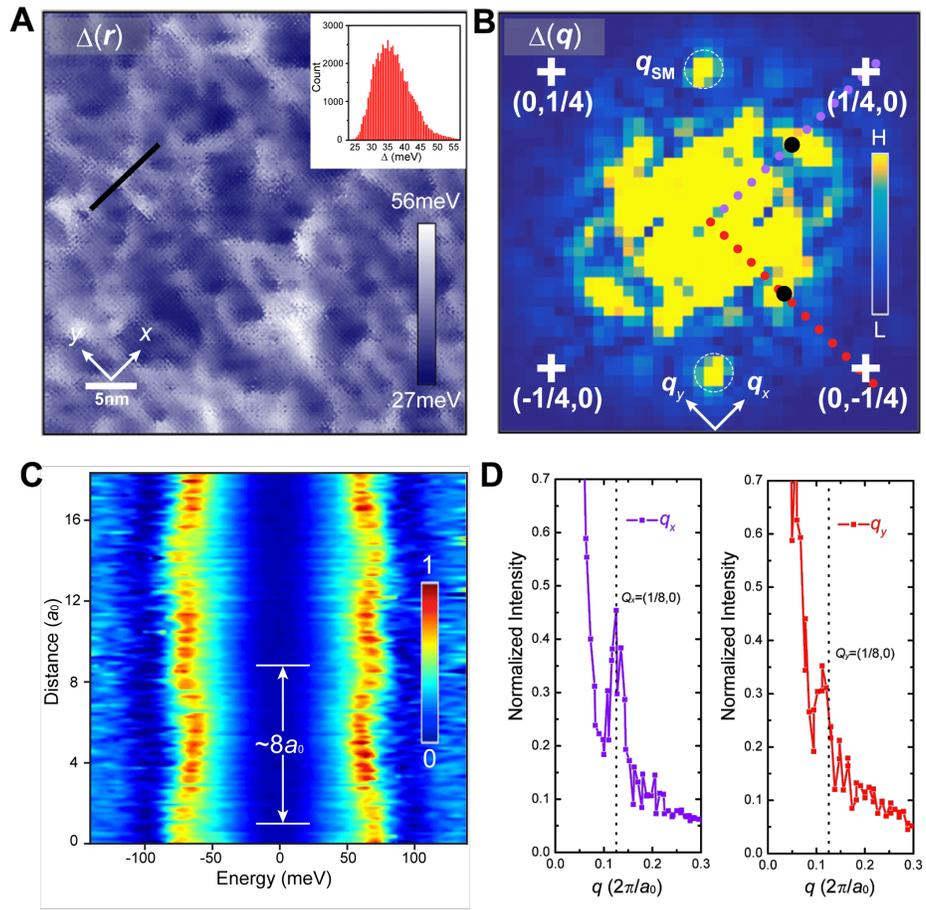

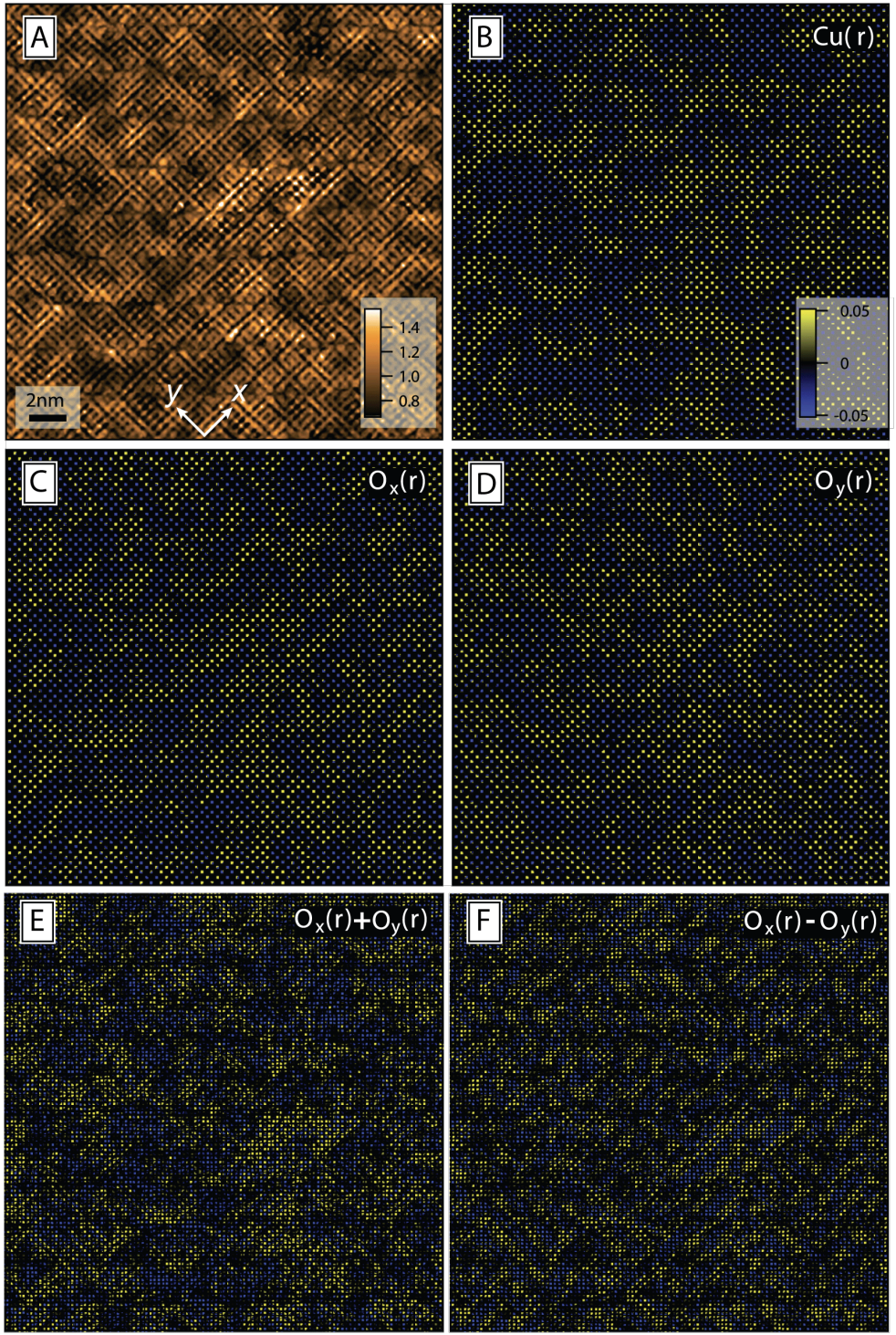

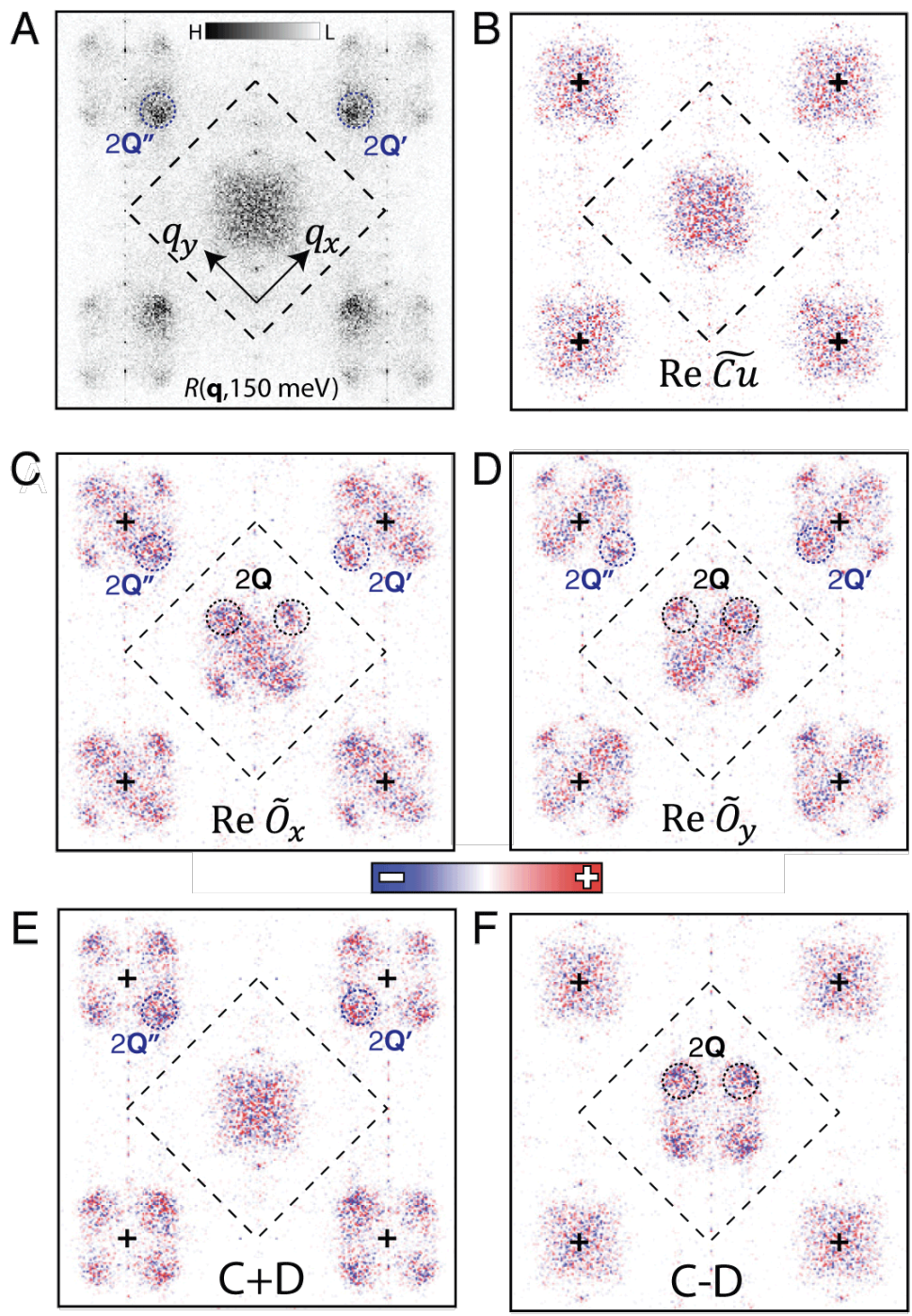

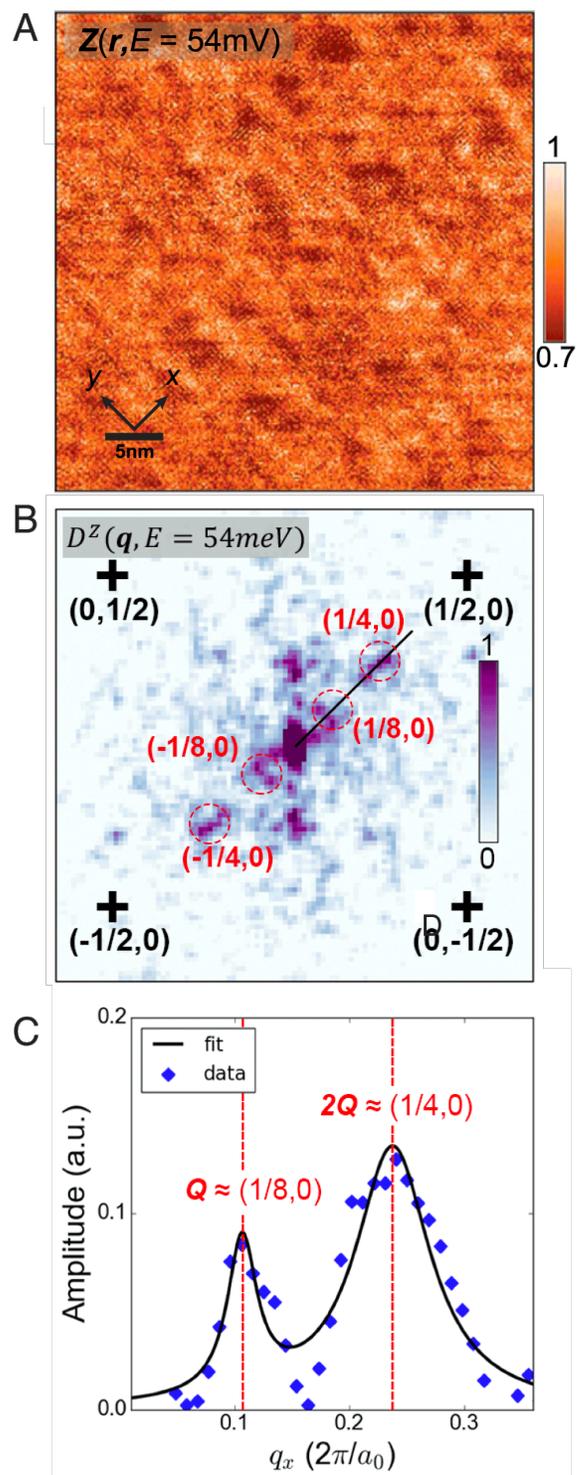

**Figure 7**

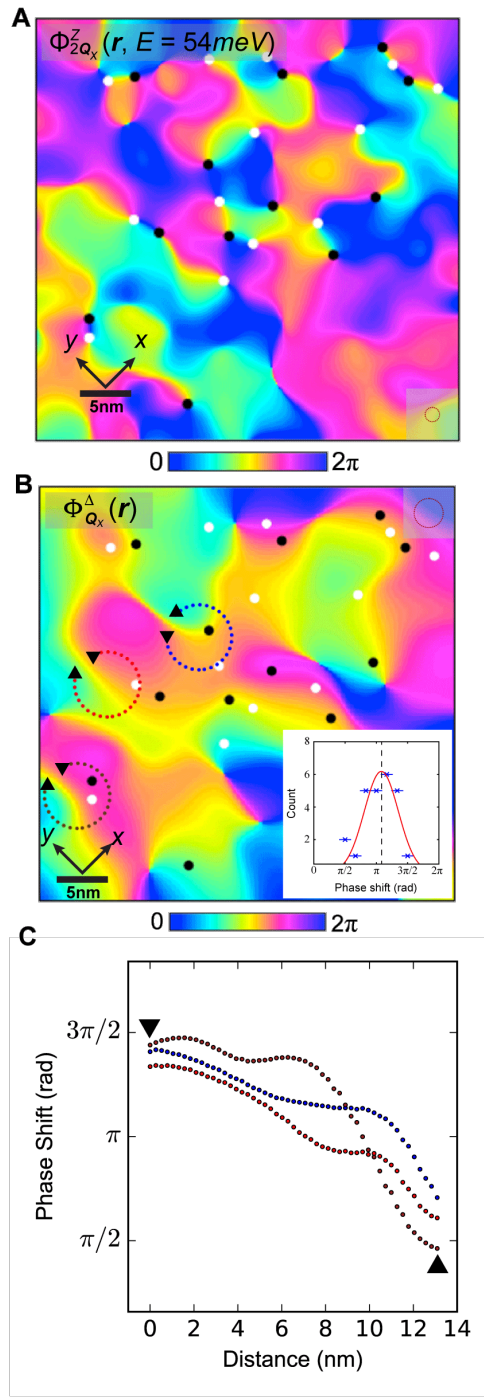

**Figure 8**

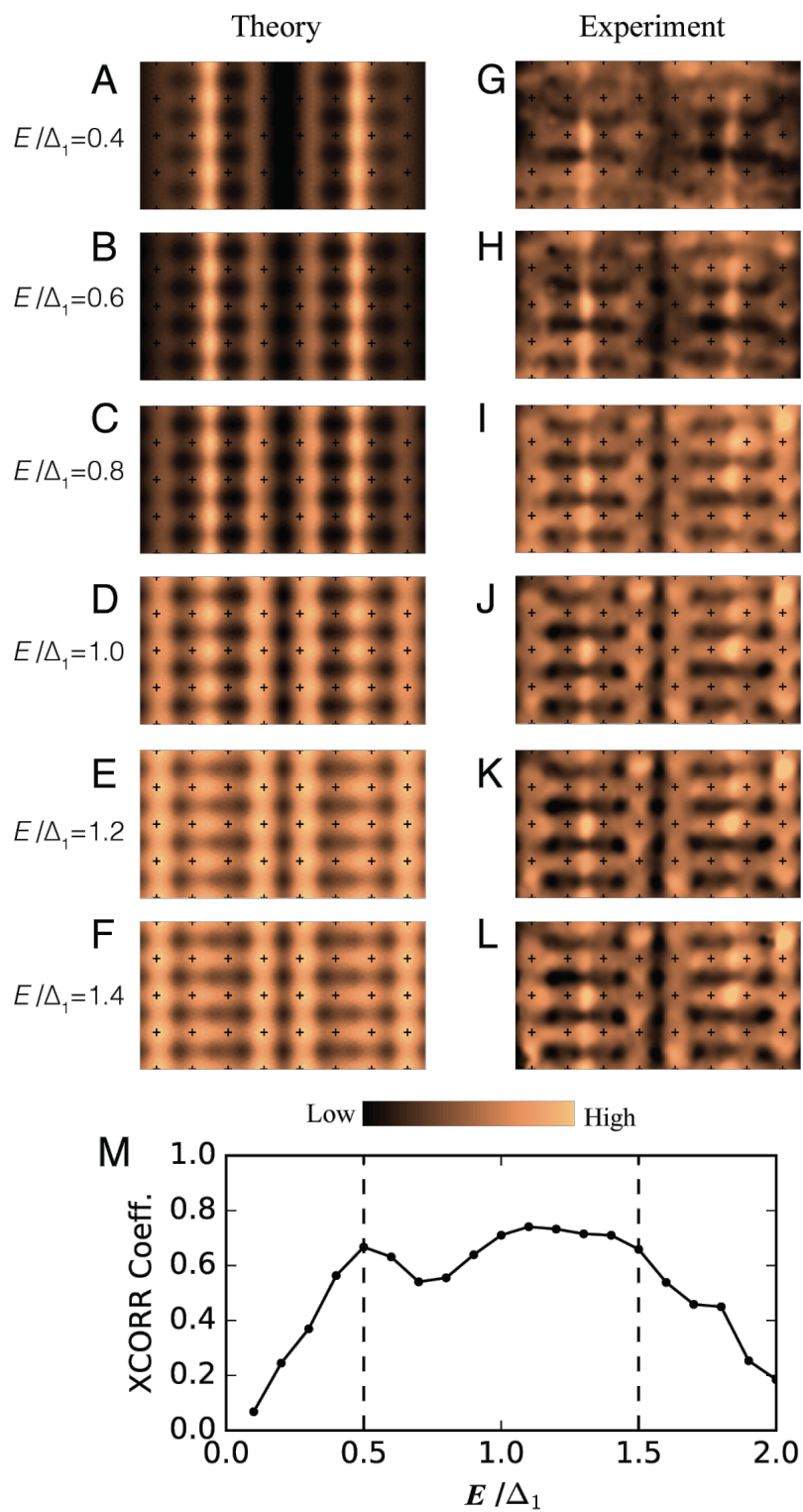

Theory                    Experiment

A   $E/\Delta_1=0.4$      G

B   $E/\Delta_1=0.6$      H

C   $E/\Delta_1=0.8$      I

D   $E/\Delta_1=1.0$      J

E   $E/\Delta_1=1.2$      K

F   $E/\Delta_1=1.4$      L

Low ████ High

M

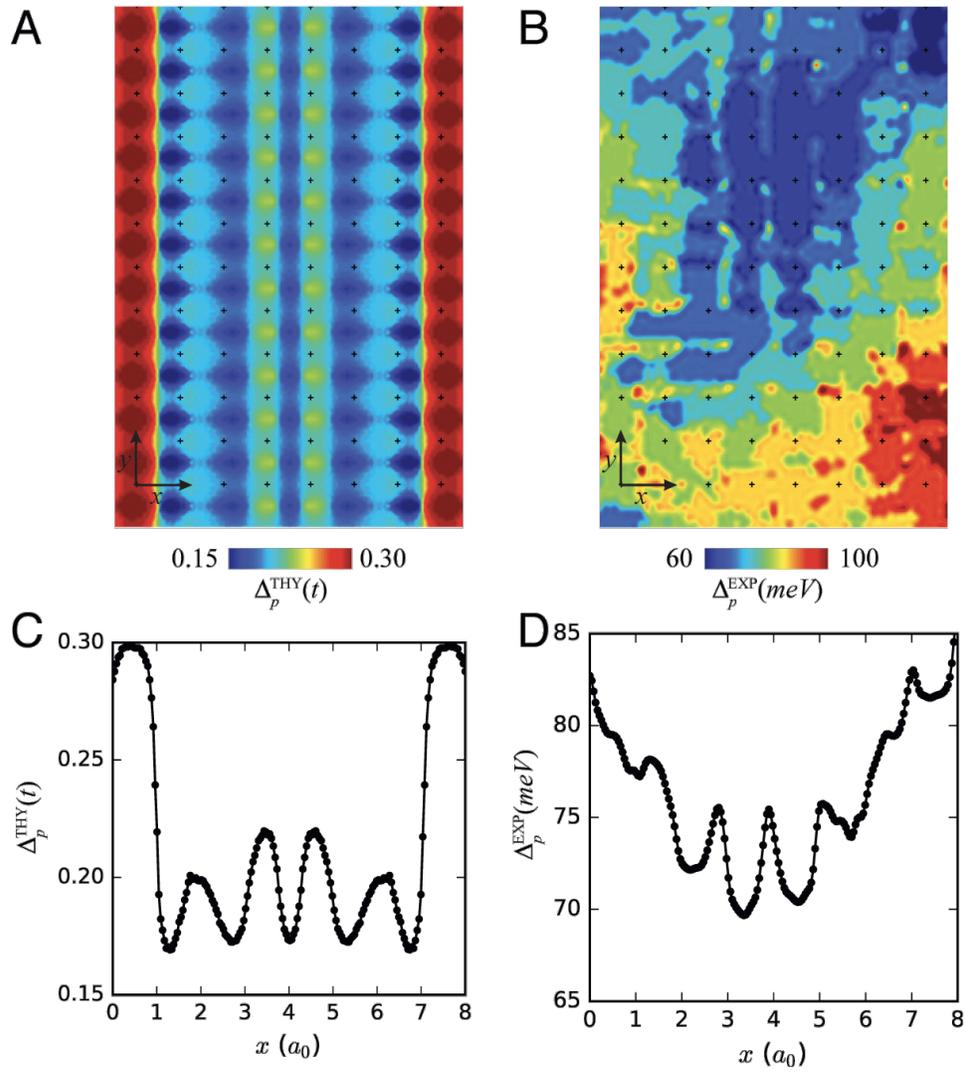

A

B

0.15 ■■■■■■■■■■ 0.30
$\Delta_p^{\mathrm{THY}}(t)$

60 ■■■■■■■■■■ 100
$\Delta_p^{\mathrm{EXP}}(meV)$

C

D